\documentclass[12pt]{article}
\usepackage{amsmath}
\usepackage{graphicx,psfrag,epsf}
\usepackage{enumerate}
\usepackage{natbib}
\usepackage{ulem}
\usepackage{wrapfig}
\usepackage[caption=false]{subfig}

\usepackage{cleveref}
\usepackage{color}
\usepackage{amssymb}
\usepackage[abs]{overpic}

\usepackage{caption}

\newif\ifgs
\gstrue 
\ifgs
\newcommand{\red}[1]{\textcolor{black}{#1}}
\newcommand{\blue}[1]{\textcolor{black}{#1}}
\definecolor{comm}{rgb}{0.0, 0.5, 0.0}

\else
\newcommand{\red}[1]{\textcolor{red}{#1}}
\newcommand{\blue}[1]{\textcolor{blue}{#1}}
\definecolor{comm}{rgb}{0.0, 0.5, 0.0}

\fi

\newif\iffigs
\figstrue
\newcommand{\bs}[1]{\boldsymbol{#1}}

\newcommand{\argmax}{\text{argmax}}

\usepackage{mathtools}

\usepackage{setspace}

\begin{document}

\bibliographystyle{apalike}

\def\spacingset#1{\renewcommand{\baselinestretch}%
{#1}\small\normalsize} \spacingset{1}
\doublespacing

 \title{\bf Multigroup discrimination based on weighted local projections}
  \author{Thomas Ortner$^1$ \thanks{This work has been partly funded by the Vienna Science and Technology Fund (WWTF) through project ICT12-010 and by the K-project DEXHELPP through COMET - Competence Centers for Excellent Technologies, supported by BMVIT, BMWFW and the province Vienna. The COMET program is administrated by FFG.} \footnote{thomas.ortner@tuwien.ac.at}, Irene Hoffmann$^1$, Peter Filzmoser$^1$, Maia Zaharieva$^2$, \\Christian Breiteneder$^2$, Sarka Brodinova$^1$ \\
    $^1$Institute of Statistics and Mathematical Methods in Economics \\
    $^2$Institute of Software Technology and Interactive Systems \\
    TU Wien}
\maketitle

% Abstract & Keywords _________________________________________________
\begin{abstract}
\blue{A novel approach for supervised classification analysis for high dimensional and flat data (more variables than observations) is proposed. We use the information of class-membership of observations to determine groups of observations locally describing the group structure. By projecting the data on the subspace spanned by those groups, local projections are defined based on the projection concepts from \cite{ortner2017guided} and \cite{ortner2017local}. For each local projection a local discriminant analysis (LDA) model is computed using the information within the projection space as well as the distance to the projection space. The models provide information about the quality of separation for each class combination. Based on this information, weights are defined for aggregating the LDA-based posterior probabilities of each subspace to a new overall probability. The same weights are used for classifying new observations.}

\blue{In addition to the provided methodology, implemented in the R-package \textit{lop}, a method of visualizing the connectivity of groups in high-dimensional spaces is proposed at the basis of the posterior probabilities. A thorough evaluation is performed using three different real-world datasets\blue{, underlining the strengths of local projection based classification and the provided visualization methodology}.}
\end{abstract}

% Introduction _________________________________________________
\section{Introduction}

Supervised classification methods are widely used in research and industry, including tasks like tumor classification, speech recognition, or the classification of food quality. Observations are gathered from $G$ distinct groups and for each observation the group membership is known. Decision boundaries are then estimated in the sample space, such that a new observation can be assigned to one of the $G$ groups. The aim of discrimination methods is to find 
\blue{classification}
boundaries, which result in low misclassification rates for new observations, i.e. new observations are assigned to the correct class with high accuracy. 

\blue{
Linear discriminant analysis (LDA) is a popular tool for classification. It estimates linear decision boundaries by maximizing the between-group to within-group variance and assumes equal covariance structure of the groups. LDA often gives surprisingly good results in low-dimensional settings, however, it cannot be directly applied if the number of variables exceeds the number of observations since then the within-group covariance estimate becomes singular and its inverse cannot be calculated. With restrictions on the covariance estimation the problem of singularity can be mended but asymptotically (with increasing \blue{number of variables}) the performance of LDA  is not better than random guessing \cite[][]{bickel2004some, shao2011sparse}.}

In many classification tasks, it is commonly the case that the underlying data has a flat structure, i.e.~there are more variables than observations. Therefore, a great variety of alternative classification methods and extensions of LDA have been developed to overcome its limitations. Several proposed approaches consider projection of the data onto a lower dimensional subspace \cite[][]{barker2003partial, chen2013near} or reducing the dimensionality by model-based variable selection \cite[][]{witten2011penalized}. 
Other methods are not based on covariance estimation and so they are not restricted to low-dimensional (non-flat) settings, e.g. k-nearest neighbour (KNN) classification, support vector machines (SVM) or random forests (RF). Nevertheless, the noise accumulation due to a large number of variables, which are not informative for the class separations, affects these methods as well. \red{Also the concepts of combining projections with classification methods has been previously explored \citep[e.g.][]{lee2005projection, caragea2001gaining} where the focus is  on exploratory classification for finding suitable projections for visualization.}

We propose a new approach for supervised classification based on a series of projections into low-dimensional subspaces, referred to as local projections. In each subspace, we calculate an LDA model. % based on the projected training observations and their distance to the projection space.
The posterior probabilities of each LDA model are aggregated (weighted by a class-specific quality measure of the projection space) to obtain a final classification. The idea of aggregating posterior probabilities in the context of random forests has been proposed by \citet[][]{bosch2007image} taking the average over the posterior probabilities from all trees.
%Each subspace represents a local description of an observation.

% Describe LDA problems \cite{hastie1995penalized}
%\cite{fan2008high} - LDA and high dim, independence assumption
%\cite{chen2013near} - PCA-LDA

\blue{
The remainder of the paper is structured as follows.  Section~\ref{sec:methodology} presents the proposed method. First, local projections based on the k-nearest class neighbors of an observation and distances within and to the projection space are introduced in Section~\ref{ss:localprojections}, resulting in the local discrimination space where an LDA model is estimated. Next, in Section~\ref{ss:weighting}, we introduce weights used for aggregating the posterior probabilities from the individual LDA models leading to a final classification rule. 
In Section~\ref{ss:trainingLDA}, the range of the tuning parameter $k$, associated with the dimensionality of the local discrimination space, is discussed and a strategy to select the tuning parameter is presented.
Section~\ref{sec:visualization} introduces a way of visualizing the data structure and the degree of separation.  In Section~\ref{sec:evaluation}, three real-world datasets are used to evaluate the performance of our approach in comparison to other related and popular classification methods. The datasets cover settings with only 25 and up to almost 10.000 variables, including multigoup and binary classification problems, and a dataset where subgroups are known to exist. The effect of imbalanced group sizes in the training data is investigated and results are visualized by the techniques introduced in the previous section. Section~\ref{sec:conclusion} concludes the paper.}

\section{Methodology}\label{sec:methodology}

Let $\bs X$ denote a data matrix of $n$ observations $\bs X = ( \bs x_1, \dots , \bs x_n)'$ in a $p$-dimensional space, $\bs x_i \in \mathbb R^p$, $i=1,\dots ,n$. We further assume the presence of $G$ classes where the class memberships of the observations are stored in a categorical vector $\bs y$ with $y_i=g$ iff $\bs x_i$ comes from group $g$, for $g\in\{1,\dots,G\}$. The number of observations in group $g$ is denoted by $n_g$ with $n=n_1 + \dots + n_G$. For all observations we assume that they have been drawn from $G$ different continuous 
\blue{probability distributions}.

\blue{For our methodology, it is important that each space spanned by a subset of $k$ observations has a dimension of at least $G-1$ and that there are no ties present in our data. These requirements automatically imply high-dimensional dataspaces as the area of application. Both assumptions can be met by a preprocessing step, removing duplicate or linearly dependent observations. Note that these restrictions only apply for the training data but not for new observations.}
%\blue{BEM: Was ist wenn Variablen linear abhaengig sind? Nimmt man dann die PCs zu den EW$>0$?}

%This assumption guarantees with a probability of one that no ties are present in the data as well as in the distances between observations. 
%Note that this assumption is not critical as in case of ties, we can perform a preprocessing step removing duplicate observations from our dataset as duplicated observations can always be assumed to be member of the same class (at least in the prediction model).

%Even though LDA is often criticized for making too strong assumptions \cite[e.g.][]{aa} (e.g. assuming the same covariance structure for each group, when being applied on an appropriate data setup), it remains one of the most efficient classification methods available even when some of the assumptions are violated. 

Previous research \citep{ortner2017guided, ortner2017local} shows the effectiveness of using series of projections to overcome the limitations caused by a flat data structure. 
In this section, the local discrimination method is introduced, which allows for the number of variables $p$ to exceed the number of observations $n$. The idea is as follows. For a fixed observation $\bs x_i$, its $k$ nearest neighbours are identified, called the \textit{core} of $\bs x_i$, which are used to define a $k-1$ dimensional hyperplane, the core space. The Euclidean distance to this hyperplane, called \textit{orthogonal distance}, is calculated for each observations. The hyperplane and the orthogonal distance together define a $k$-dimensional subspace, the local discrimination space, where an LDA model is estimated. This approach is performed for each observation resulting in $n$ LDA models. To assign the class membership to an observation, its posterior probabilities of all models are aggregated.

\subsection{Local discrimination space}
\label{ss:localprojections}

Let $d^g_k(\bs x)$ denote the $k$th-smallest distance from $\bs x$ to any observation from class $g$, for $g\in\{1,\dots,G\}$. 
According to \cite{ortner2017guided} and \cite{ortner2017local}, we define the \textit{core} of $\bs x_i$ as the k-nearest class neighbors of $\bs x_i$,
\begin{equation}
core(\bs x_i) = \{ \bs x_j: d( \bs x_i, \bs x_j ) \leq d_k^g(\bs x_i) \wedge y_i = y_j = g   \} = \{ \bs x_{i_1}, \dots , \bs x_{i_k} \},
\end{equation}
where $d( \bs x_i, \bs x_j ) $ denotes the Euclidean distance between $\bs x_i$ and $\bs x_j$, and $i_1,\dots,i_k$ are the indices of the core observations within $\bs X$. In contrast to \cite{ortner2017local}, we use all k-nearest class neighbours as we can use the group membership in order to guarantee a \textit{clean} core, i.e. no observations from other groups within the core. 

Any of the $n$ available cores $core(\bs x_1), \dots, core(\bs x_n)$ can be used to unambiguously define an affine subspace spanned by the core observations. In order to determine the projection onto this subspace, we center and scale the data with respect to the $k$ core observations $\bs x_{i_j}$, $j=1,\dots , k$.
\begin{align}
\hat{\bs \mu}_{i}  = & \frac{1}{ k} \sum_{j=1}^k \bs x_{i_j}  \label{eq:center} \\
\bs{\hat{ \sigma}}_{i} =& \left( \sqrt{\hat{Var}(x_{i_11}, \dots, x_{i_k1})}, \dots, 
 \sqrt{\hat{Var}(x_{i_1p}, \dots, x_{i_kp})} \right)' \\ 
\nonumber
= & (\hat{\sigma}_{i1}, \dots, \hat{\sigma}_{ip})' ,
\end{align}
where $\hat{Var}$ denotes the sample variance. For the ongoing work, we denote $\tilde{\bs X}^i = (\tilde{\bs x}_{1}^i,\dots,\tilde{\bs x}_{n}^i)'$ as the data matrix of centered and scaled observations
based on the location and scale estimators $\hat{\bs \mu}_i$ and $\hat{\bs \sigma}_i$ of the core of $\bs x_i$.
A projection onto the subspace spanned by the core of $\bs x_i$ is defined by $\bs  V_i$ from the singular value decomposition (SVD) of the centered and scaled core observations $(\tilde{\bs x}_{i_1}^i,\dots,\tilde{\bs x}_{i_k}^i)'=\bs U_i \bs D_i \bs V_i'$. Since the core of $\bs x_i$ consists of exactly $k$ linearly independent observations, $\bs D_i$ is a $k-1$ dimensional diagonal matrix with non-zero singular values in the diagonal.  

Since the idea of no ties being present in the data and each core consisting of linearly independent observations may appear like a strong limitation, an adjustment of the definitions can help \blue{in order to avoid a preprocessing step}. If we interpret the core of $\bs x_i$ as a set of observations, where iteratively the observation from the same class, closest to $\bs x_i$ is added until a $k-1$ dimensional subspace is spanned, we only need to guarantee the existence of such cores, which is a much weaker assumption. 

Given the projection matrix $\bs V_i$ from the decomposition, a  representation of the data $\bs X$ in the core space is defined by down-projecting the centered and scaled data matrix, $\bs Z^i =  \tilde{\bs X}^i \bs V_i$. The core representation consists of $k-1$ orthogonal variables, while the $p-k$ dimensional complement of $\bs Z^i$ defines the orthogonal complement of the core space. In contrast to commonly used procedures of first reducing the dimensionality using PCA and then performing a discrimination method like LDA, we acknowledge the fact that the last principal components might contain an important part of the information like exploited by modern outlier detection algorithms~\cite[e.g.][]{hubert2005robpca, kriegel2012outlier}. Since the reduction of dimensionality remains vital, we aggregate the information from the orthogonal complement by considering the Euclidean distance to the core space,
\begin{equation}
OD^i(\bs x_j) = ||  \tilde{\bs x}^i_j  -  \bs V_i\bs z^i_j  ||,
\end{equation}
where $\bs z^i_j = \bs V'_i \tilde{\bs x}^i_j$ denotes the core representation of $\bs x_j$ given $core(\bs x_i)$. 

The combination of the core representation and the orthogonal distance $OD$ in a matrix, $[\bs Z^i, OD^i]$, provides a $k$-dimensional representation for all observations of $\bs X$. \blue{This $k$-dimensional space is the \textit{local discrimination space}. The reduction of the sample space to the local discrimination space results in a good description of the neighbourhood of an observation $\bs x_i$ and also includes grouping structure which is not described in the core space by the orthogonal distances.}

\blue{An LDA model is estimated in the local discrimination space, excluding the observations from the core of $\bs x_i$. It is necesarry to exclude the core observations, because they have very specific properties in the local discrimination space and this would distort the within-group covariance estimation. In \cite{ortner2017guided} it is shown that 
\begin{align}
OD^i(\bs x_j) &= 0 &  \forall \bs x_j \in core(\bs x_i) \label{eq:coreOD}\\
SD^i(\bs x_j) &\equiv const. &   \forall \bs x_j \in core(\bs x_i),  \label{eq:coreSD}
\end{align}
\blue{where SD represents the score distance defined as the Euclidean distance within the core space.}
These properties hold because for $\bs x_j \in core(\bs x_i)$ the full information is located in the core space, so the orthogonal distances are zero. The scaling applied to the data based on the covariance estimation of the core observations leads to constant score distances for $\bs x_j \in core(\bs x_i)$. So the core observations must not be included in the computation of the LDA model. The model estimated on the remaining observations in the local discrimination space is denoted by $LDA_i$.}

\blue{For the model $LDA_i$ the posterior probability of group $g$ given an observation $\bs x$  is defined by
\begin{align}\label{eq:posteriorprob}
P_{LDA_i}(g \mid \bs x)=\frac{h_g(\bs x) }{\sum_{j=1}^G h_l(\bs x) } ,
\end{align}
where $h_g(\bs x)$ denotes the estimated density of a multivariate normal distribution with the group mean of class $g$ as center and the pooled within-group covariance matrix as covariance estimate.}

\subsection{Weighting/aggregating local projections} \label{ss:weighting}

We now have a set of $n$ local discrimination spaces and their respective LDA models. In order to receive an overall classification rule for a new observation $\bs x$, we need to aggregate the $n$ available models from the core spaces. \blue{ We accomplish such an aggregation by using the posterior probabilities defined in Equation \eqref{eq:posteriorprob}. First we consider the mean over all $n$ posterior probabilities of $\bs x$ belonging to group $g$, for $g\in\{1,\dots,G\}$,}
\begin{equation}
\tilde{P}_{LP^k_1}(g \mid \bs x )  = \frac{1}{n} \sum_{j=1}^n P_{LDA_j}(g \mid \bs x)  \label{eq:postLP_const1} ,
\end{equation}
and we define the aggregated posterior probability of $\bs x$ belonging to group $g$, for $g\in\{1,\dots,G\}$, as
\begin{equation}
P_{LP^k_1}(g \mid \bs x)  = \frac{ \tilde{P}_{LP^k_1}(g \mid \bs x) }{ \sum_{j=1}^G \tilde{P}_{LP^k_1}(j \mid \bs x) } \label{eq:postLP_const2} .
\end{equation}
These new aggregated posterior probabilities are based on a fixed number $k$ describing the number of core observations as indicated by the index of $LP_1^k$. 

The posterior probabilities of the LDA models, $P_{LDA_i}(g \mid \bs x)$, compared to the true class membership of $\bs x$ reflect the quality of separation in the respective local projection. We distinguish between two quality measures. \blue{Let $q^{g+}_{i}$ denote the mean posterior probability of belonging to class $g$ over all observations actually coming from class $g$, with respect to the model $LDA_i$, i.e.
\begin{equation}
q^{g+}_i = \frac{1}{n_g} \sum_{k: y_k = g} P_{LDA_i}(g \mid \bs x_k )
\end{equation}
and $q_i^{g-}$ the mean posterior probability of non-class-$g$ observations being classified as class $g$ observations given the model $LDA_i$, i.e.}
\begin{equation}
q_i^{g-} = \frac{1}{n - n_g} \sum_{k: y_k \neq g} P_{LDA_i}(g \mid \bs x_k ).
\end{equation}
Based on $q_i^{g+}$ and $q_i^{g-}$, we define weights $w_i^g$ representing the quality of each local projection $i=1, \dots,n$ for each group $g\in\{1,\dots,G\}$,
\begin{equation}
w_i^g = exp\left( q_i^{g+} - q_i^{g-} \right) .
\end{equation}

Based on these quality measures $w_i^g$, we redefine the overall posterior probabilities from Equation \eqref{eq:postLP_const2} by weighting each projection for each class with the respective weight. Note that these weights are class-specific and, therefore, a class-individual standardization of weights is required. In our notation, we remove the subscript $1$ from Equation~\eqref{eq:postLP_const1} and Equation~\eqref{eq:postLP_const2}, which represents constant weights of $1$ for each local projection, resulting in:
\begin{align}
\tilde{P}_{LP^k}(g \mid \bs x ) & = \frac{1}{\sum_{i=1}^n w_i^g}  \sum_{i=1}^n w_i^g P_{LDA_i}(g \mid \bs x ) \\
P_{LP^k}(g \mid \bs x ) & = \frac{ \tilde{P}_{LP^k}(g \mid \bs x ) }{ \sum_{j=1}^G \tilde{P}_{LP^k}(j \mid \bs x ) } \label{eq:pp}
\end{align}
Equivalently to classical LDA, we use these posterior probabilities to assign an observation $\bs x$ to a class $\hat{y} = \argmax_{g\in\{1,\dots,G\}} P_{LP^k}(g \mid \bs x )$. \blue{This decision rule defines the local discrimination model.}

\subsection{The choice of $k$}
%Training discriminant models on local projections and limitations on the choice of $k$}
\label{ss:trainingLDA}

\blue{The computation of LDA models in the full dimensional space, given more variables than observations are available, requires data preprocessing including dimension reduction \cite[e.g.][]{barker2003partial, chen2013near} or the parallel performance of model estimation and variable selection \cite[e.g.][]{witten2011penalized, hoffmann2016sparse}. The concept of local projections allows us to compute an LDA model for each local projection due to the low dimensional core space. The parameter determining the dimensionality is $k$ of the $k$-nearest class neighbours. It is important to properly tune $k$ since it defines the degree of locality for each projection. Smaller $k$'s are able to better describe a lower dimensional manifold on which groups might be located but increase the risk of  not being able to properly describe the local data structure.}

%While training discrimination models in the full dimensional space poses \blue{various} problems and demands for complex approaches of parallel dimension reduction \citep[e.g.][]{Irene2017}, the local projection space as defined defined in Section \ref{ss:localprojections} allows to directly train the model due to the lower dimensionality. The choice of the configuration parameter $k$ of the $k$-nearest class neighbours is important since it defines the degree of locality for each projection. Smaller $k$'s are able to regularize a lower dimensional manifold on which groups might be located but increases the risk of  not being able to properly describe the local data structure.

\blue{The number of classes $G$ as well as the number of observations $n_g$ for $g\in\{1,\dots,G\}$ provide a first limitation for the range of $k$. In order to compute an LDA model with G classes, a dimensionality equal to at least $G-1$ is required. Therefore,
\begin{equation}
G-1 \leq k
\end{equation}
provides a lower boundary for $k$.}

\blue{To identify an upper boundary for $k$, two properties of the core observations must be taken into account.
Due to the specific properties of the core observations stated in Equations \eqref{eq:coreOD} and \eqref{eq:coreSD}, they are not included in the computation of the LDA model. Therefore, an upper boundary for $k$ is given by 
\begin{equation}
k \leq n-(k+1)
\end{equation}
to guarantee a non-singular covariance estimation. It is useful to further reduce the upper boundary of $k$ in order to allow for a reasonable covariance estimation. Here we take three times more observations than variables leading to the limitation
\begin{equation}
3k \leq n-k.
\end{equation}}

\blue{With these restrictions on $k$, LDA models in the core spaces can be computed but for the evaluation of the models further limitations are necessary. To be able to evaluate the LDA models, we depend on the posterior probabilities of observations for each class in order to determine the risks of misclassification. Since a core consists of observations from the same class only and the core observations are excluded from the LDA model, the size of the smallest class needs to exceed $k$. 
\begin{equation}
k+1 \leq \min_{g\in \{ 1, \dots, G \}} n_g-1 
\end{equation}}
\blue{
Due to the identified restrictions on $k$, we optimize $k$ within the following interval:
\begin{equation}\label{eq:intervalk}
\left[ G-1, \min\left( \frac{n}{4} , \min_{g \in \{ 1, \dots, G \}} n_g -1  \right) \right]
\end{equation}}

\blue{For a given $k$, the misclassification rate of the local discrimination model is calculated by summing up the number of misclassified observations (again excluding the core observations) divided by $n-k$, the total number of observations. The tuning parameter $k$ is chosen from within the interval described in Equation \eqref{eq:intervalk} in such a way that the misclassification rate is minimized. }

\section{Visualization of the discrimination}
\label{sec:visualization}
%\label{sec:evaluation}

\blue{In linear discriminant analysis, the projection space is used for the visualization of the discrimination \cite[e.g.][]{hair1998multivariate}. The Mahalanobis distances of observations to the class centers refer to the posterior probabilities of the observations for the respective classes. This approach is not feasible for local discrimination since each LDA model refers to a different subspace and the aggregated posterior probabilities do not refer to one specific low dimensional space, where the posterior probabilities could be visualized.}

%In contrast to visualization of linear discriminant analysis, where the projection space used for the discrimination can be visualized \cite[e.g.][]{hair1998multivariate}, our classification is based on an aggregation of $n$ projections and the corresponding LDA models. Therefore, a low-dimensional representation of the aggregated models can not be directly determined. % from the subspaces and the quality measures. 

We \blue{therefore} focus on visualizing the aggregated posterior probabilities and follow an approach for compositional data using ternary diagrams. We present the visualization technique on the four-group \textit{Olitos} dataset which is used as a benchmark dataset for robust, high-dimensional data analysis. The dataset is publicly available in the R-package \textit{rrcovHD} and was originally described by \cite{armanino1989chemometric}. 

\cite{hron2013robust} used ternary diagrams to visualize the outcome of \blue{(three-group)} fuzzy clustering results, which can be interpreted the same way as posterior probabilities of discrimination models.
The difficult aspect about ternary diagrams is the limitation to three variables. Therefore, we select two classes, use the respective \blue{posterior} probabilities and as third composition the sum of \blue{posterior} probabilities for all remaining classes. \blue{This new three-class composition is visualized in Figure \ref{fig:ternary1}.}

\iffigs
\begin{figure}[!htb]
\centering
\includegraphics[width=0.4\linewidth]{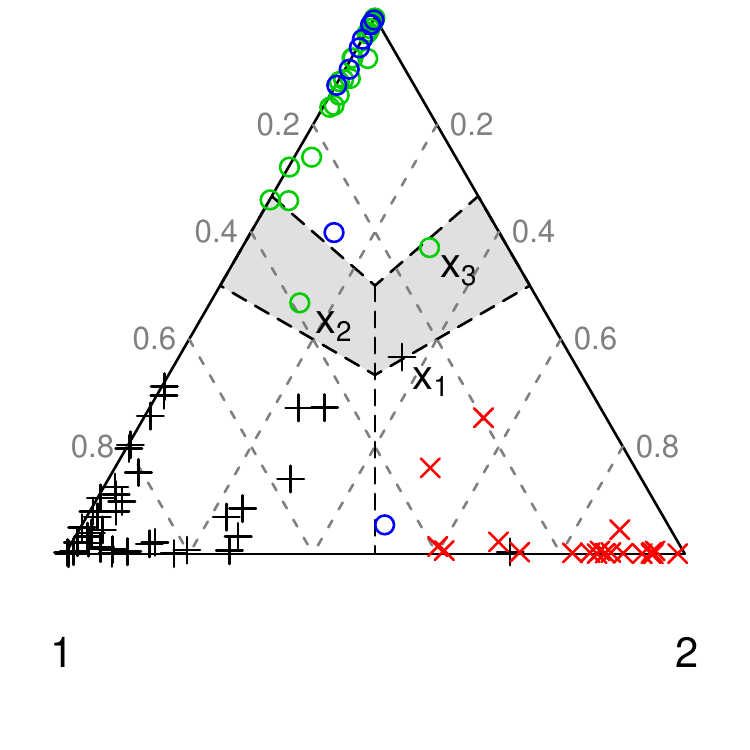} 
\caption{The aggregated posterior probabilities of two classes (class 1 and class 2) compared to the remaining classes is visualized as a ternary diagram. The dashed lines represent classification rules. Observations located in the white areas can be assigned to the respective group, while the grey area represents an uncertain area, where no reliable statement can be made. The grey dashed lines refer to posterior probabilities of the selected classes.}
\label{fig:ternary1}
\end{figure}
\fi

\blue{Figure \ref{fig:ternary1} shows the proposed representation for a sample of the Olitos dataset. The focus of this representation is the evaluation of the separation between the two selected classes $1$ and $2$. The gray dashed lines and the numbers on the left side show the posterior probability for an observation to belong to group 1.
The two white areas at the bottom separated by a vertical dashed line represent the classification rules for the separation between class 1 and 2. Observations in the left area are assigned to group 1 and in the right area to group 2. In the bottom right area we can identify one outlier from the \textbf{blue} class and one from class $1$  which are wrongly assigned to group 2. Besides these two false classifications, additional information can be gained from the diagram.}

\blue{First, the grey area represents the region, where no statement about classification can be made with certainty. Two observations $x_1$ and $x_2$ are highlighted there. While $x_1$ is located in the uncertainty area, we can still tell that it will be misclassified since the posterior probability for class $2$ is larger than for class $1$. This decision is indicated by the vertical dashed line within the uncertainty area. However, from this figure, it is not possible to say whether or not it will be assigned to class 2 or to one of the other classes. \blue{The same holds for observation $x_2$.
The posterior probability for class $1$ is close to 0.4, for class $2$ to 0.15. Therefore, the posterior probabilities for classes $3$ and $4$ sum up to approximately 0.45. 
Depending on the class-specific allocation, the maximal posterior probability for the classes $3$ and $4$ varies between 0.225 and 0.45 and the largest posterior probability for $x_2$ can originate from class $1$, $3$ or $4$.}
% can be used to visualize any classification.  
% Depending on the number of aggregated classes, the properties of this diagram changes. We }
}
% \blue{BEM: Zahlen stimmen oben nicht:
% Therefore, the posterior probabilities for classes $3$ and $4$ sum up to approximately 0.4. Thus, the largest posterior probability for this observation can be approximately 0.4, and it can originate from class 1, 3 or 4.}

\blue{Second, the white classification area at the top of the triangle visualizes those observations, which with certainty will not be assigned to class 1 or class 2. We note a minor risk of misclassifying observations in the direction of class $1$. Note that the size of the uncertainty area and therefore the size of the third classification area highly depends on the number of groups to be aggregated. In a 3-group case, all posterior probabilities can be visualized and no area of uncertainty exists as shown in Figure \ref{fig:ternary2a}. The remaining plots of Figure \ref{fig:ternary2} show the impact of increasing numbers of groups on the area of uncertainty.}

\iffigs
\begin{figure}[!htb]
\centering
 
\subfloat[]{
\label{fig:ternary2a} 
    \includegraphics[width=0.18\linewidth]{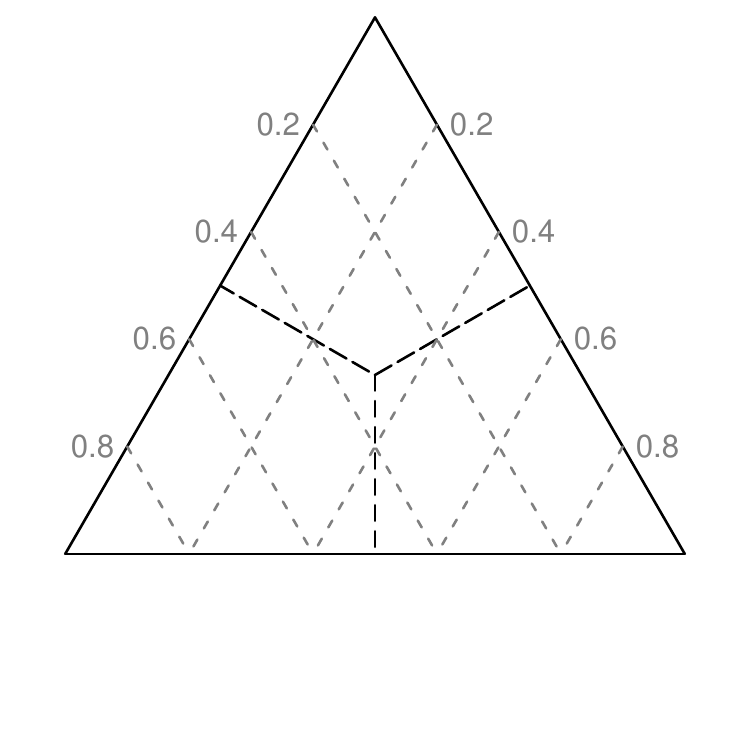} }
\subfloat[]{
    \label{fig:ternary2b}
    \includegraphics[width=0.18\linewidth]{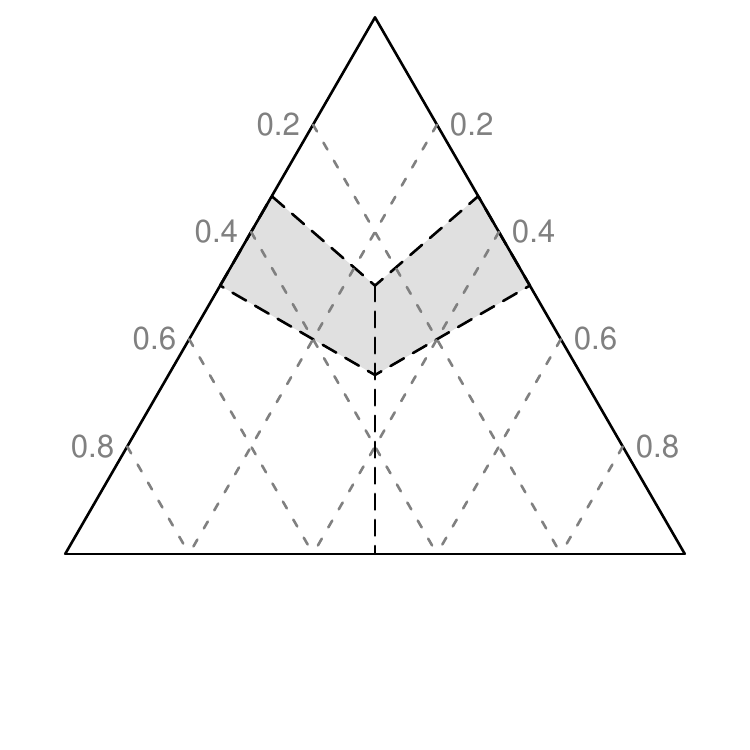} }
\subfloat[]{
    \label{fig:ternary2c}
    \includegraphics[width=0.18\linewidth]{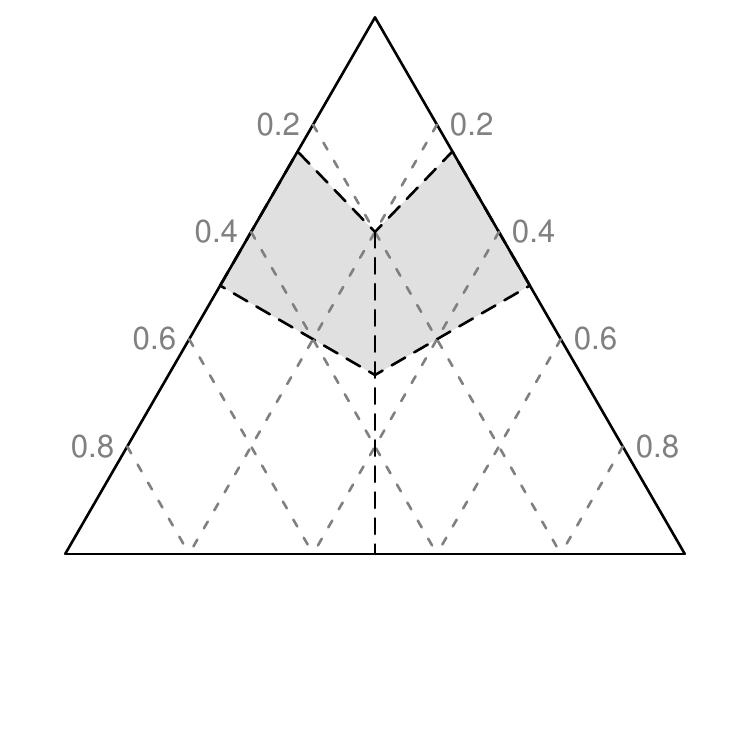} }
\subfloat[]{
    \label{fig:ternary2d}
    \includegraphics[width=0.18\linewidth]{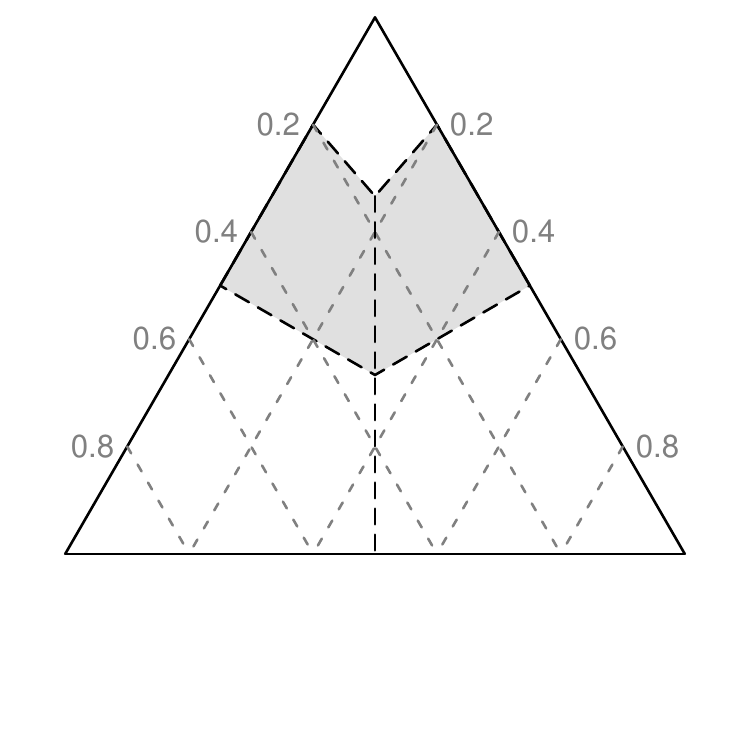} }
\subfloat[]{
    \label{fig:ternary2e}
    \includegraphics[width=0.18\linewidth]{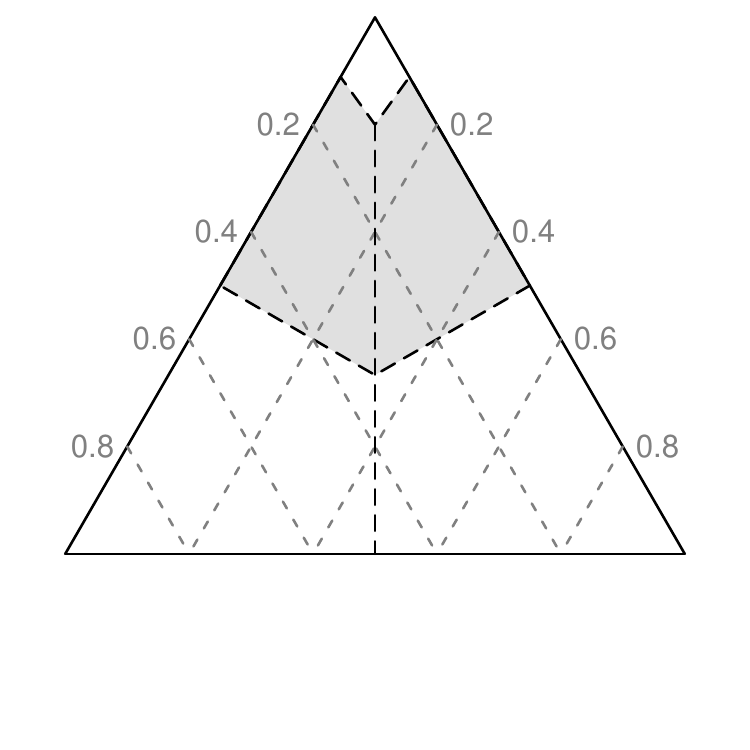} }

\caption{The effect of the number of aggregated groups is visualized. Figure (a) refers to a three-class case, (b) to a four-class case, (c) to a five-class case, (d) to a six-class case and (e) to a ten-class case.}
\label{fig:ternary2}
\end{figure}
\fi

\blue{Finally, the positioning of the observations in the ternary diagram provides some insight on the connection between the groups. The red observations from class $2$ in Figure \ref{fig:ternary1} are mostly aligned along the axis from $1$ to $2$. Observations from the aggregated group are also mostly aligned between $1$ and $3$ while the black observations split up between class $2$ and $3$. Therefore, class $2$ is strongly connected to class $1$ but has no connection to the aggregated classes. Observations like $x_3$ strongly deviate from the typical class direction and should therefore be candidates for further investigation in the context of outlier analysis.}

\iffigs
\begin{figure}[!thb]
\centering
\includegraphics[width=0.75 \linewidth]{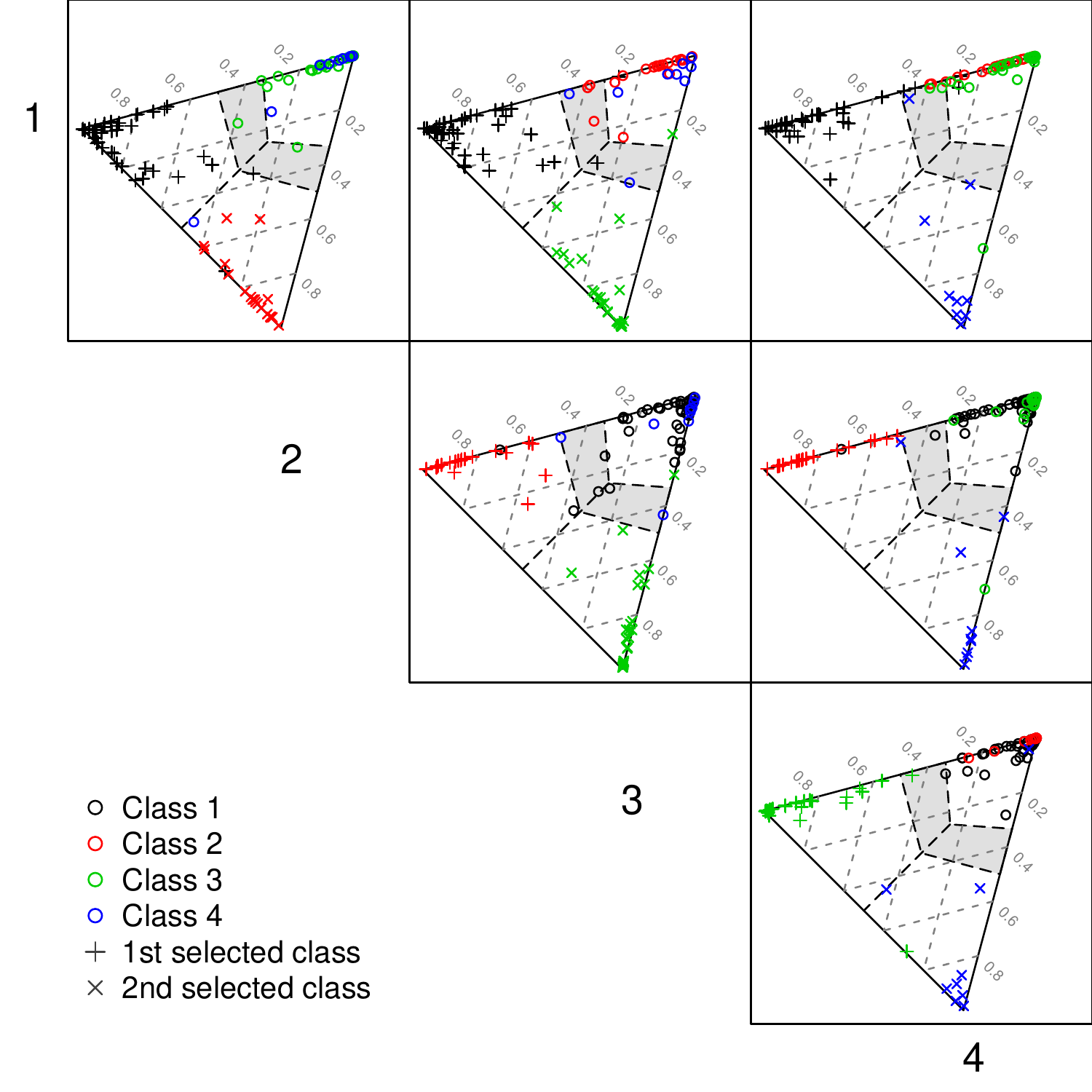} 
\caption{A set of ternary diagrams is used to visualize the classification performance for each possible combination of classes. While the color remains constant, we switch labels to emphasize the currently selected classes labeled on the bottom and left of the diagrams.}
\label{fig:ternarymat}
\end{figure}
\fi

% The discrimination model visualized in Figure \ref{fig:ternary} provides a fairly good description. We note one blue observation located between class 1 and 2 in the red classification area, where the probabilities of class membership drastically deviates from the ground truth. In addition two black observation are misclassified and a few observations are located close to the classification rule. All other relevant observations (observations from the selected classes) are classified correctly with the vast majority of observations being located beyond the $80\%$ posterior probability.

Since the representation in Figure \ref{fig:ternary1} uses a \blue{three-components} representation, which can not illustrate the overall discrimination result, we propose to use a combination of ternary diagrams in the form of a \blue{scatterplot matrix}, as presented in Figure \ref{fig:ternarymat}. Each combination of possible \blue{two-class plus one aggregated group} classifications is presented as described for Figure \ref{fig:ternary1}. In order to align the groups and increase the readability, the diagrams have been rotated accordingly. 

Besides providing information on the quality of discrimination and risks of misclassification, we can derive an \blue{overall picture about group connectivity. We already remarked on the location of class $2$. The positioning of the observations of the green and blue observations in the second and third column of the matrix reveals that the green group has stronger ties to the black group, while the blue observations are equally drawn to its direct neighbors, the black and the blue group. Such insight on connectivity provides a feeling for the location of groups in high dimensional spaces which is in general a non-trivial task limited by the human spacial sense.}

%information about group specific separation. E.g. in combination $2-4$ we note that besides one outlying blue observations, class 2 is very well separated from class 4. This becomes evident as the observations line up along the sidelines to the third corner meaning that the risk of misclassification emerges from connections to the remaining groups but a high probability of class 2 never occurs simultaneously with a high probability for class 4 and vice verse. Combination $1-2$ provides a strong contrast showing that class 1 and class 2 are located close to each other. Considering the alignment of observations in the ternary plots provides a good insight on the connections between classes and the overall positioning of classes in the sample space. The advantages of this approach over classical visualization of discrimination models becomes evident in \blue{the next section} where we compare the different approaches on real-world datasets.

\section{Evaluation}
\label{sec:evaluation}

In order to evaluate the performance of our proposed local discrimination approach\blue{, abbreviated by LP for Local Projections,} we use three real-world datasets which have previously been used as benchmark datasets for high dimensional data analysis. Based on those datasets, we compare with well-established classification methods from the fields of computer science and statistics. While the visualization introduced in Section \ref{sec:visualization} provides interesting insights into each dataset and will be provided as well, we focus on comparing the used methods based on the misclassification rate. 

For each dataset, we split the available observations into training and test dataset. The same training dataset is used for each  method to estimate the discrimination model and the same test dataset is used to evaluate the performance of all the models by reporting the misclassification rates. 

The employed datasets consist of groups of different numbers of observations. 
\blue{Since the outcome of a method can be strongly affected by the specific choice of the training and test set,}
we resample the observations 50 times per dataset, creating a series of training and test datasets resulting in a series of misclassification rates. The overall performance is then measured based on the median misclassification rate as well as on the deviation from the median misclassification rate.

\subsection{Compared methods}

The selection of classification methods is based on the popularity of the methods, the importance for our setups, and the relevance for our proposed approach. The most important aspect is the applicability on the evaluated datasets. The crucial factor is the flat data structure (more variables than observations), especially in class-specific subsets of the overall dataset. In order to cover related classification methods, we include Linear Discrimination Analysis (LDA) as this is the classification method internally used for each local projection. We further include statistical advancements of LDA, which try to deal with disadvantageous properties of our datasets of interest, namely penalized LDA and partial least squares for discriminant analysis. The most related method from the field of computer science is KNN-classification as our local projections are based on a knn-estimation. The last methods included in the evaluation are support vector machines and random forests to cover the most commonly used classification approaches from the field of computer science.

For \textbf{Linear Discriminant Analysis} (LDA), it is assumed that the covariance structure is the same for each class and has elliptical shape. Under this assumption, the optimal decision boundaries to separate the groups are linear. The separation of the classes is achieved by taking $G-1$ orthogonal directions which maximize the within-group variance to the between-group variance. In this $G-1$ dimensional space, the Euclidean distance to the group centers is used to assign an observation to the group with the closest center.

For the calculations, the \texttt{lda} function from the R-package \texttt{MASS} is used. This implementation can be applied to data with $p>n$ by performing singular value decomposition and reducing the dimensionality to the rank of the data.

\textbf{Penalized LDA} (PLDA) introduced by \citet[][]{witten2011penalized} is a regularized version of Fisher's linear discriminant analysis. A penalty on the discriminant vectors favours zero entries, which leads to variable selection. The influence of the penalty is controlled by the sparsity parameter $\lambda$: larger values of $\lambda$ lead to fewer variables in the model.

The sparsity parameter $\lambda$ is selected from 10 values between $10^{-4}$ and $5$ by 10-fold cross-validation on the training data using as selection criterion the minimum mean misclassification rate. The number of discriminating vectors is set to $G-1$. The functions for cross validation and model estimation are provided in the R-package \texttt{penalizedLDA} \citep{peanlizedLDA}.

\textbf{Partial least squares for discriminant analysis} (PLSDA) was theoretically established by \cite{barker2003partial}, where its relationship to LDA and the application to flat data was discussed. PLSDA performs in a first step a projection onto $K$ latent variables, which considers the grouping information of $y$. Then LDA is performed in the reduced space.

For the evaluation the R-package \texttt{DiscriMiner} \citep{DiscriMiner} is used, which provides code for the selection of the number of components $K$ by leave-one-out cross-validation. 
%(no explanation how $K$ is selected, code not so clear for me... needs more time to investigate)
%\blue{Vielleicht auch keine weitere Erklaerung noetig.}

\textbf{Support Vector Machines} (SVMs) are a popular machine learning method for classification. 
\blue{The margins between the groups of the training data are maximized in a data space induced by the selected kernel.}
While a variety of kernels is available (e.g. linear, polynomial, sigmoid, etc.), we limit the optimization procedure to the radial basis kernel, which is suggested as standard configuration.

We use an R-interface to \textit{libsvm} \citep{chang2011libsvm} included in the R-package \textit{e1071}. The internal optimization of SVM is based on a $k$-fold cross-validation on the training dataset\blue{, providing a range of values for the cost parameter and for $\gamma$}. 
For multi-class-classification, libsvm internally trains K(K-1)/2 binary ‘one-against-one’ classifiers based on a sparse data representation matrix.

\textbf{Random Forest} (RF) is an ensemble-based learning method commonly used for classification and regression tasks. It builds a forest of decision trees using bootstrap samples of the training data and random feature selection for each tree. The final prediction is made as an average or majority vote of the predictions of the ensemble of all trees.

The RF implementation in the R-package \textit{randomForest} uses Breiman's random forest algorithm \citep{breiman2001random} for multigroup classification. In order to optimize the classification model, we use the internal optimization procedure starting with $\sqrt{p}$ randomly sampled variables as candidates for splits and increase this number with a factor of 1.5 in each optimization step.  

In \textbf{KNN-Classification} (KNN), the class-membership of the $k$-nearest neighbors of an observation based on Euclidean distances is used for determining the class of the respective observation. For $k=1$,
the class of the nearest neighbor is used, for
$k>1$, the class with the highest frequency
is used.  In the case of ties, a random decision is performed.  We use one-fold cross-validation in order to optimize $k$ individually for each sampled dataset.

\subsection{Olive oil}

%We want to consider datasets with various specific features. 
The first dataset in our experiments consists of 120 samples of 25 chemical compositions (fatty acids, sterols, triterpenic alcohols) of olive oils from Tuscany, Italy, and was first introduced by \citet{armanino1989chemometric}. The dataset is publicly available in the R-package \textit{rrcovHD} \citep{todorov2014rrcovhd} where it is used as a reference dataset for robust high-dimensional data analysis. 

The olive oils are separated in four classes of 50, 25, 34 and 11 observations. In order to have enough training observations from each group available, we use $80\%$ of observations for the training dataset and the remaining $20\%$ as test observations. We repeatedly create such an evaluation setup 50 times. Hence, each training dataset consists of 96 observations, which yields the only setup where we have more observations than variables available. Therefore, classical LDA is expected to perform fairly well. Note that the smallest number of training observations per class is still much smaller than the number of overall variables. Therefore, class-specific covariance estimation as it is performed in quadratic discriminant analysis \citep{friedman1989regularized} cannot be performed in this setup or on any other of our considered datasets.

LP and LDA perform exactly the same, which can be seen in Figure \ref{fig:performanceOlitos}. PLDA slightly outperforms LP, while PLSDA, SVM, RF, and especially KNN get outperformed. In most cases all variables are included in the PLDA model but only a subset of variables contributes to each discriminant vector. This variable selection leads to a slight improvement over LDA and LP.

\iffigs
\begin{figure}[!htb]
\centering
\includegraphics[width=0.7 \linewidth]{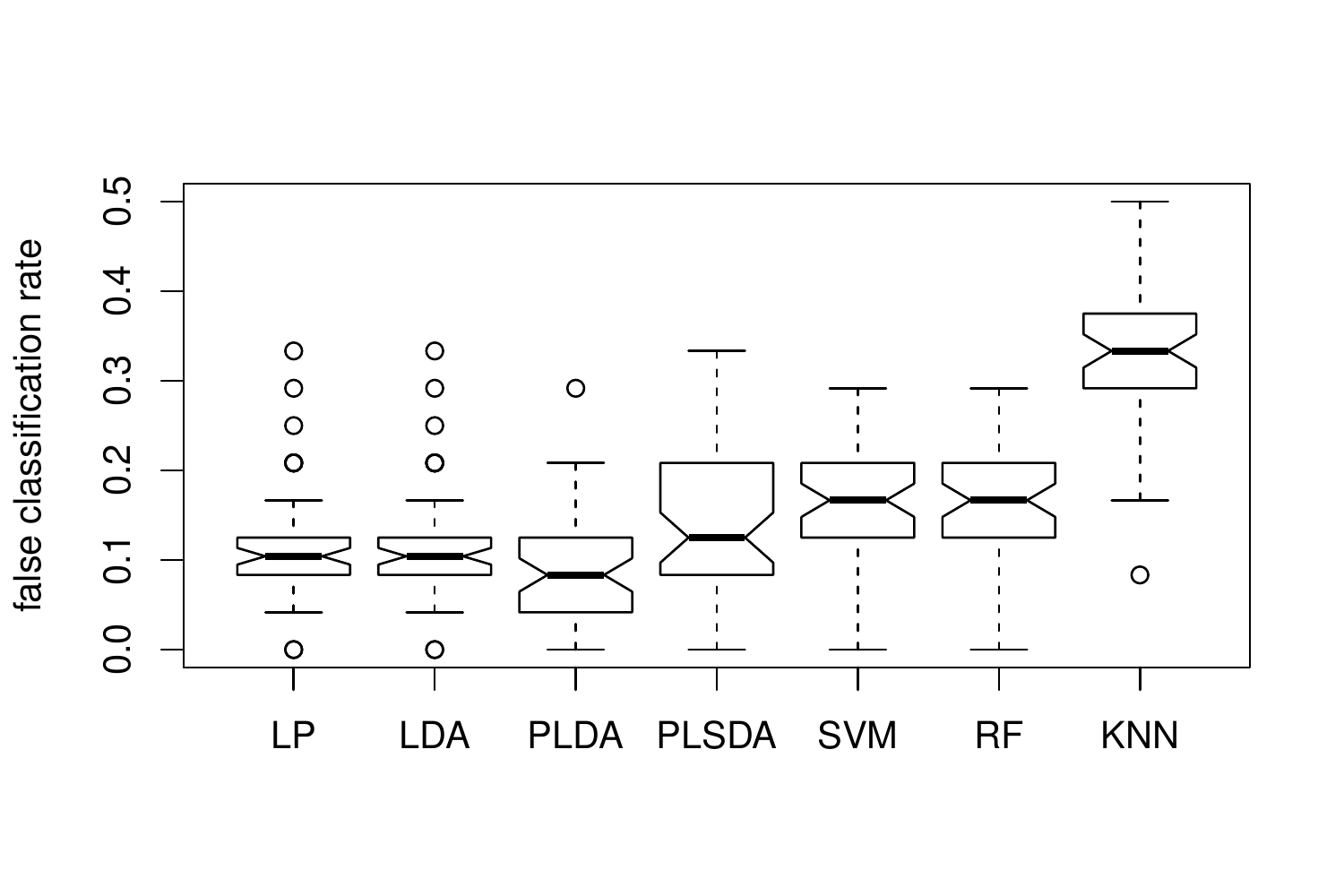} 
\caption{The performance in terms of false classification rates of all considered classification methods for 50 repetitions of the \textit{Olitos} dataset is visualized by boxplots.}
\label{fig:performanceOlitos}
\end{figure}
\fi

\subsection{Arcene}

The second real-world dataset is part of the NIPS (Neural Information Processing Systems) 2003 feature selection challenge \citep{guyon2007competitive}. The task is to distinguish between cancer and non-cancer patterns from mass-spectrometric data  with $p=9961$ variables. Therefore, we deal with a two-class separation with continuous variables. The data was obtained from two different data sources, the National Cancer Institute (NCI) and the Eastern Virginia Medical School (EVMS). The observations represent patients with ovarian or prostate cancer and health or control patients. Very small and large masses have been removed from the spectrometric data in order to compress the data. In addition, a preprocessing step including baseline removal, smoothing and scaling was performed. All these details are described in \cite{guyon2007competitive}. 

The initial setup contained of 100 training and 100 validation observations, consisting of a total of 112 non-cancer samples and 88 cancer samples. In order to have a non-equal ratio of observations \blue{to create again an imbalanced scenario}, we merge both groups and resample 22 cancer training observations and 84 non-cancer trainining observations. The remaining observations are used as test observations. This procedure is repeated 50 times as for the other datasets. 

\iffigs
\begin{figure}[!htb]
\centering
\includegraphics[width=0.7 \linewidth]{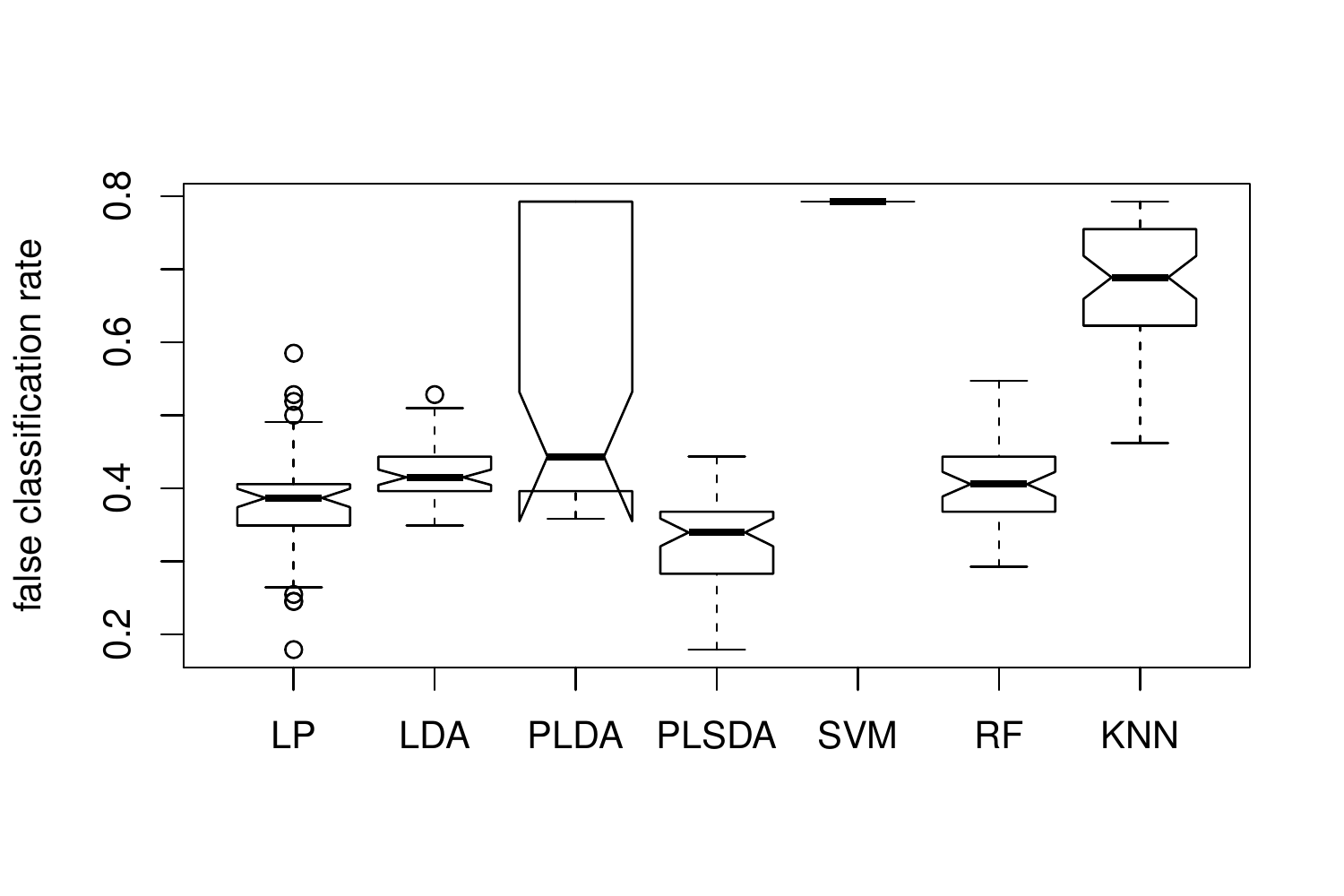} 
\caption{The performance in terms of false classification rates of all considered classification methods for 50 repetitions of the \textit{Arcene} dataset is visualized by boxplots.}
\label{fig:performanceArcene}
\end{figure}
\fi

The performance of Arcene is evaluated in terms of boxplots in Figure \ref{fig:performanceArcene}. The classification for this dataset and the designed setup is more challenging than for the other real-world datasets. LP performs well in comparison to the other evaluated approaches being outperformed only by PLSDA. The performance of 80\% false classification rate by SVM might be misleading as it appears worse than random classification. All observations from the non-cancer samples are classified as cancer samples.
\blue{This could be improved by strategies like oversampling or by changing the majority class assignment to a weighted class assignment. For LP it is not necessary to make adjustments for group imbalance.}

\subsection{Melon}

Our final dataset consists of measurements of three types of melons based on spectra analyses of 256 frequencies. The fruits are pertain to three different melon cultivars with group sizes of 490, 106 and 499 but additional subgroups are known to be present due to changes in the illumination system during the cultivation. The dataset is regularly used as a benchmark dataset for high-dimensional and robust data analysis methods \citep[e.g.][]{hubert2004fast}. Especially the subgroups usually affect non-robust analysis methods. Figure \ref{fig:datasetupMelon} provides some insight on the structure of the dataset. 

\iffigs
\begin{figure}[!thb]
\centering
\includegraphics[width=0.45 \linewidth]{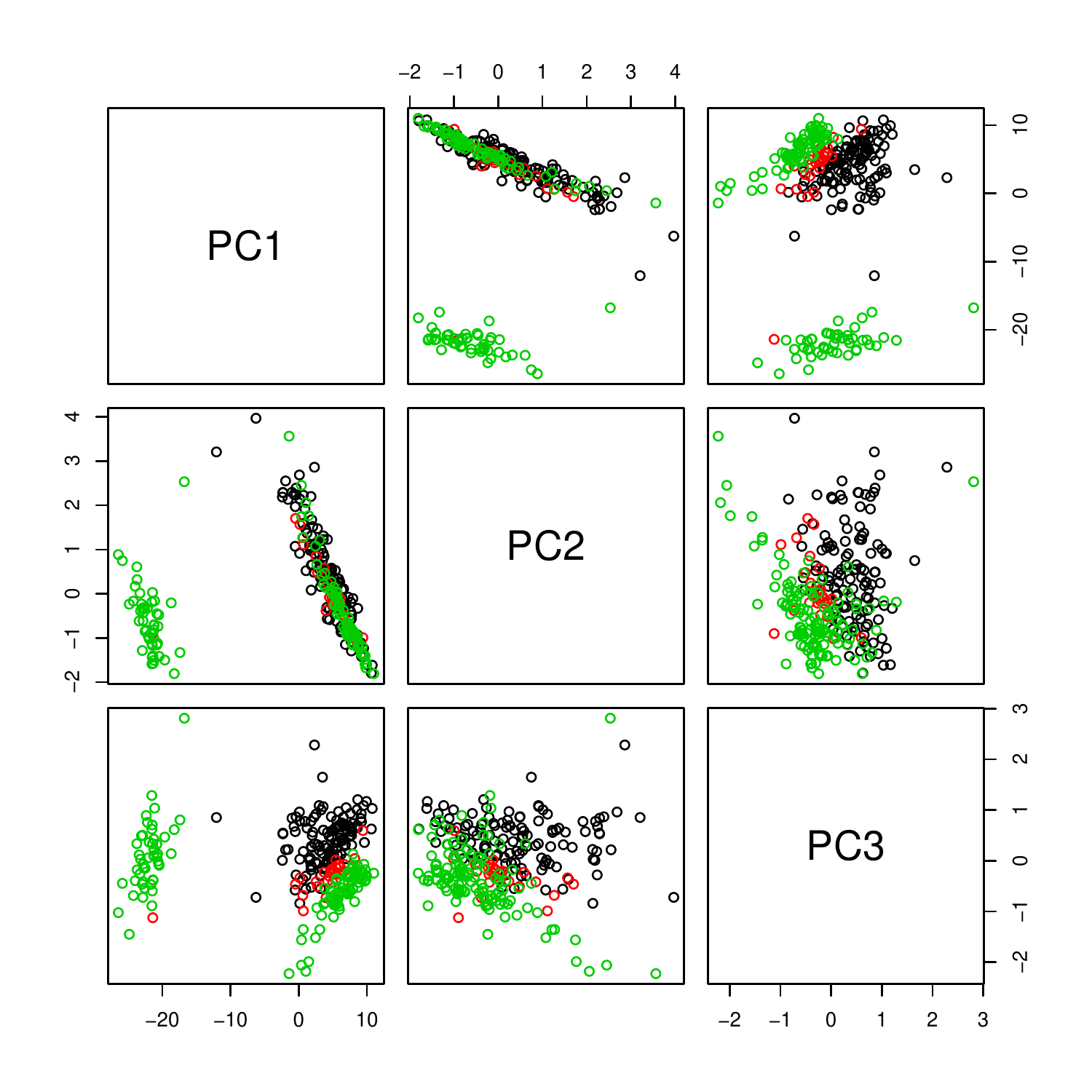} 
\caption{Visualization of the first three principal components of one sample of training observations of the \textit{Melon} dataset. We see one subgroup of the green class, represented in the first principal components and a strong overlapping structure for the remaining observations.}
\label{fig:datasetupMelon}
\end{figure}
\fi

We repeatedly sample 25\% of the observations \blue{from each group} as training observations using the remaining 75\% for testing the model performance. The smallest training class therefore consists of 26 observations leading to a complex classification problem. The performance of the compared methods is presented in Figure \ref{fig:performanceFruit}. LP can handle the challenges of the Melon dataset the best and significantly outperforms all compared methods. \blue{Especially PLDA results in a high false classification rate which is assumed to be related to the subgroups and outliers affecting the variable selection.}

\iffigs
\begin{figure}[!htb]
\centering
\includegraphics[width=0.6 \linewidth]{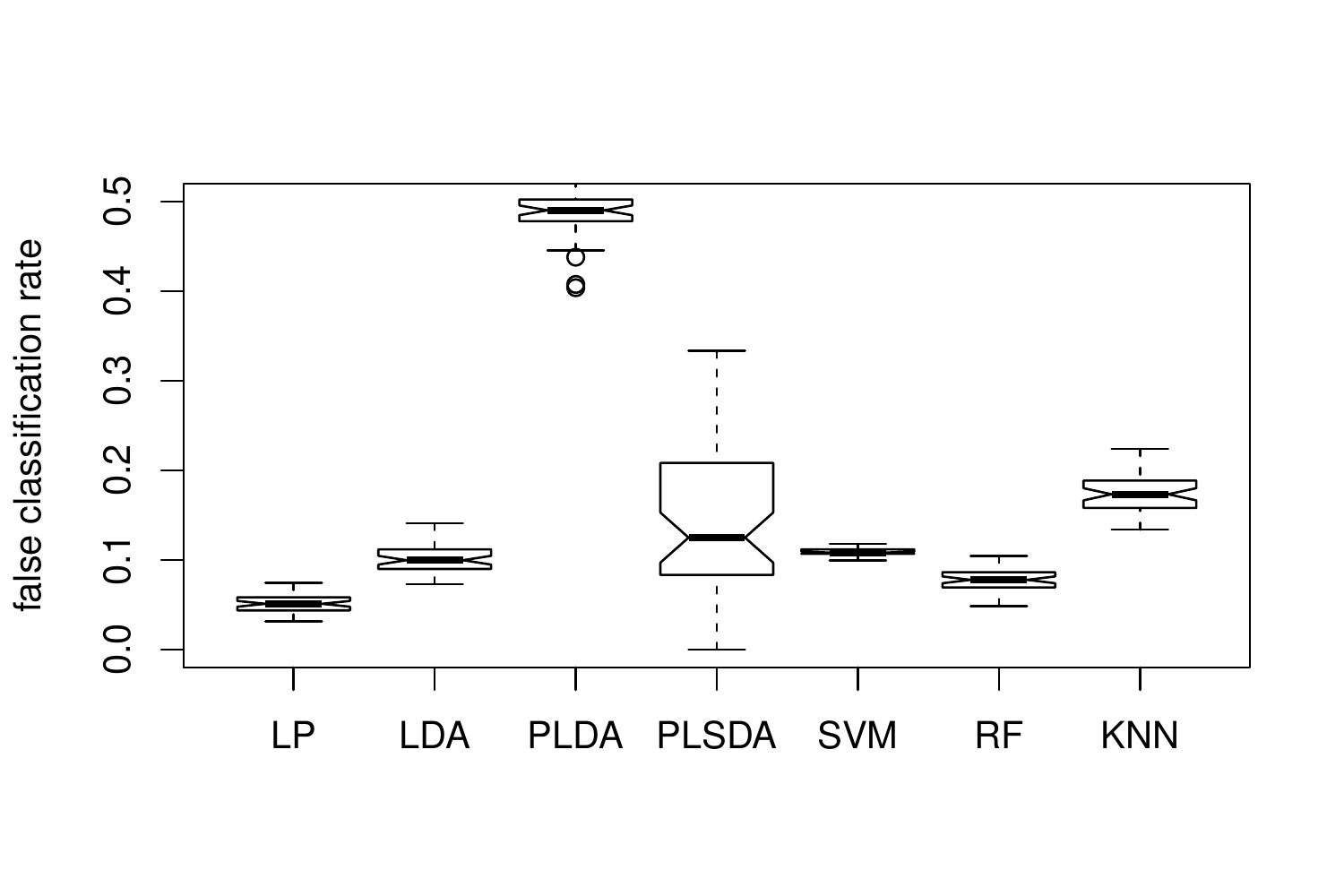} 
\caption{The performance in terms of false classification rates of all considered classification methods for 50 repetitions of the \textit{Melon} dataset is visualized by boxplots.}
\label{fig:performanceFruit}
\end{figure}
\fi

One problem during the visualization of LDA models is the property that in high-dimensional spaces with more variables than observations, the training observations will almost always be well-separated. Therefore, in a situation where we do not have enough observations to validate the model based on additional observations, a visualization of the discrimination space does not provide a lot of insight on the risks for misclassification of this model. These challenges are visualized in Figure \ref{fig:ldamodela} and Figure \ref{fig:ldamodelb}.

\iffigs
\begin{figure}[!htb]
\centering
\subfloat[]{
    \includegraphics[width=0.4\linewidth]{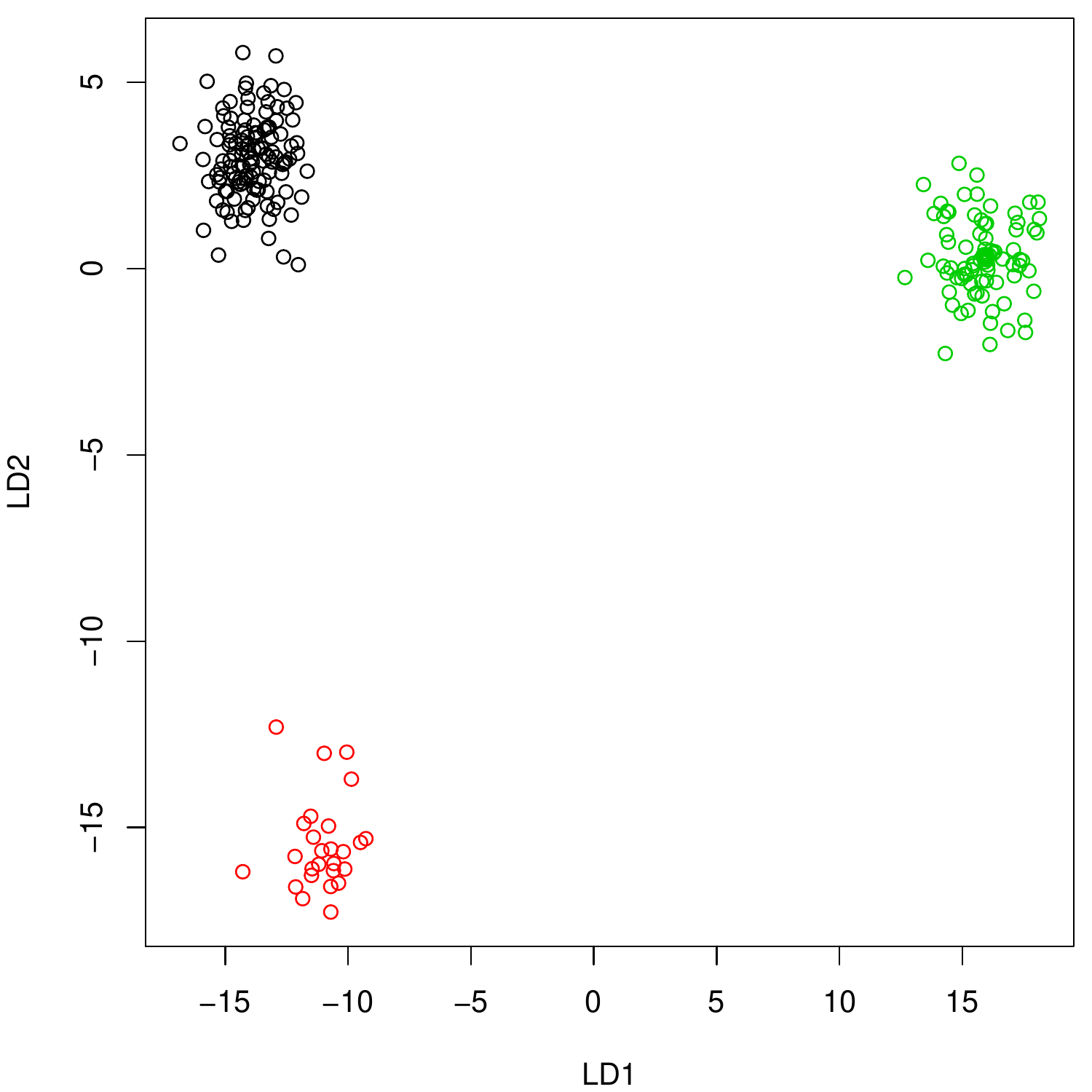} \label{fig:ldamodela} }
\subfloat[]{
    \includegraphics[width=0.4\linewidth]{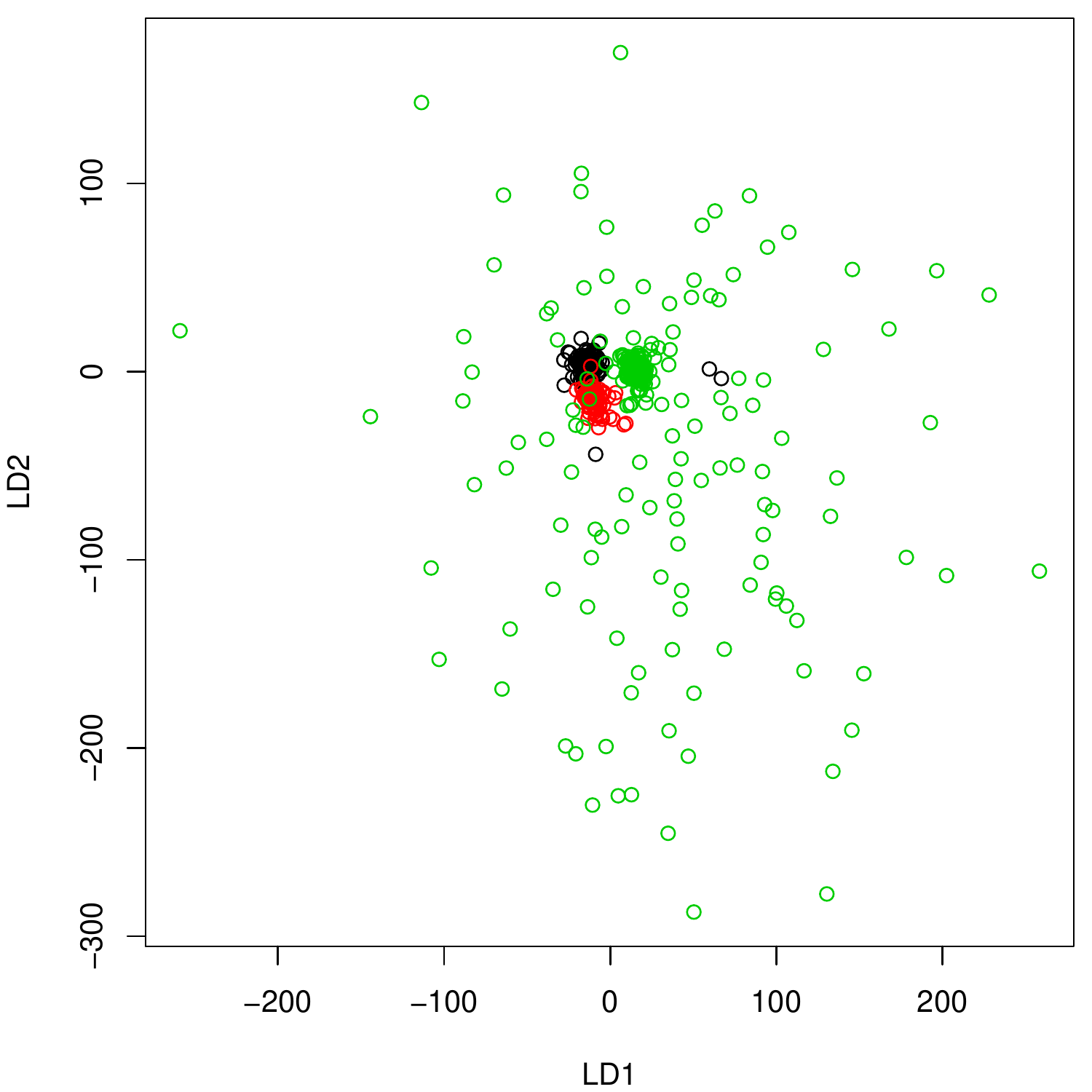} \label{fig:ldamodelb} }
\caption{Plot (a) shows the training observations of the LDA-projection space for one repetition of the \textit{Melon} evaluation. Plot (b) shows the same projection for the respective setup.}
\label{fig:visFruitLDA}
\end{figure}
\fi

We see a perfect separation in Figure \ref{fig:ldamodela} and have no indication of risks of misclassification. 
%\blue{Folgenden Satz loeschen, ist relativ kompliziert: This information can not be obtained directly from the training observations as for the training observations the posterior probabilities will be very close to $0$ or $1$, not providing additional insight without cross-validation which prevents a visual representation in a common discrimination space. BIS HIERHER.} 
\blue{The risk of misclassification can be evaluated using the
aggregated posterior probabilities of LP, defined in Equation \eqref{eq:pp}, which provide the advantage that each of the classification models is located in a low dimensional space.} Figure \ref{fig:visFruitLP} provides the visualization of the same data setup as used in Figure \ref{fig:visFruitLDA}. The risk of misclassifying observations from class $1$ as class $2$ and vice verse becomes evident in Figure \ref{fig:lpmodela} and the realization of this risk becomes evident in Figure \ref{fig:lpmodelb}. Note that this visualization can be adapted and used for posterior probabilities computed through cross-validation by any arbitrary classification method. 

\iffigs
\begin{figure}[!htb]
\centering
\subfloat[]{
    \includegraphics[width=0.45\linewidth]{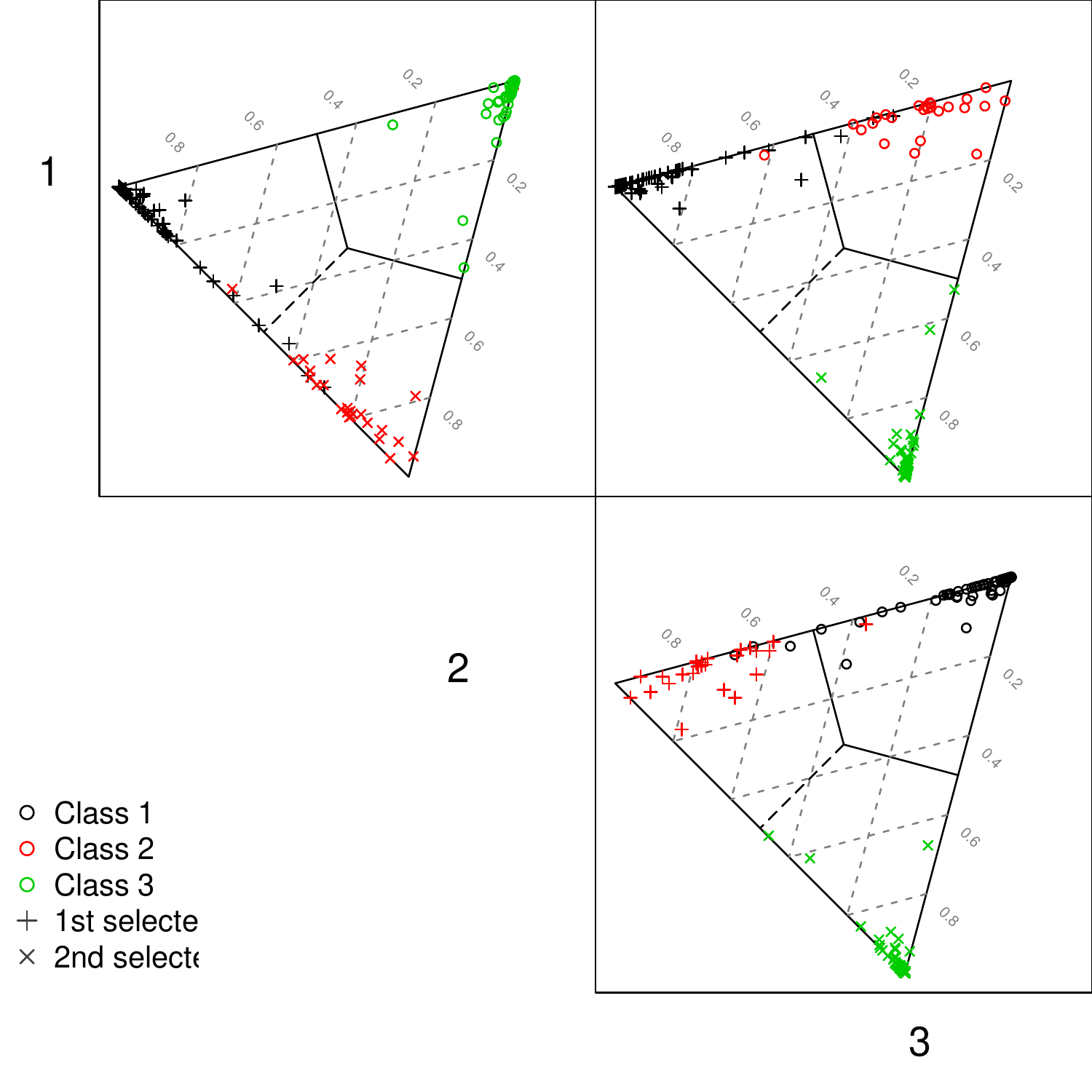} \label{fig:lpmodela} }
\subfloat[]{
    \includegraphics[width=0.45\linewidth]{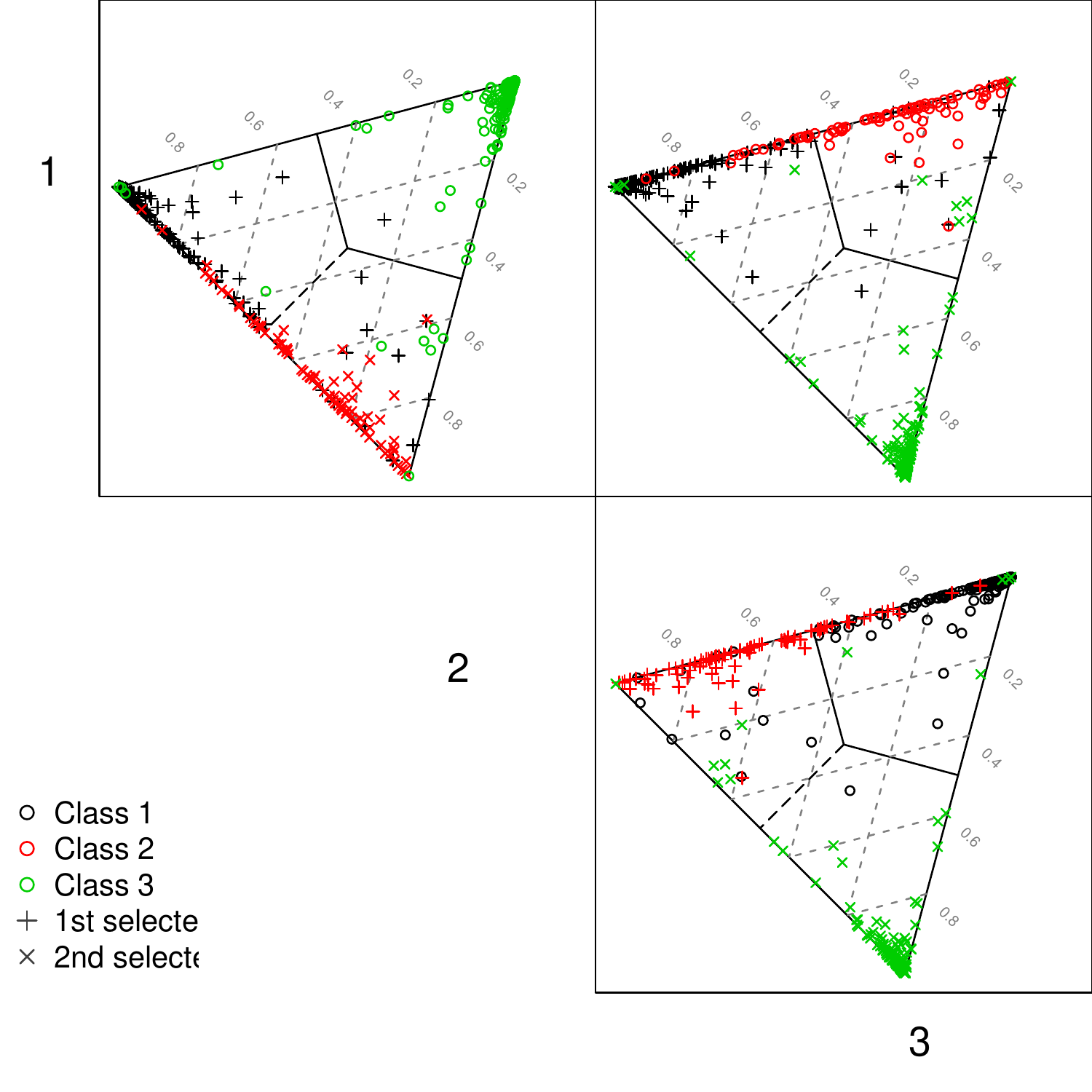} \label{fig:lpmodelb} }
\caption{The same data setup as in Figure \ref{fig:visFruitLDA} is used. Plot (a) shows the proposed visualization of aggregated posterior probabilities from local projections for the training observations. Plot (b) visualizes the same aggregations for the respective test observations.}
\label{fig:visFruitLP}
\end{figure}
\fi

\bigskip
\blue{A further experiment is carried out with the Melon dataset: While in the previous experiment, the training datasets contained 25\% of the observations from the original groups (with sizes 490, 106 and 499), we now investigate the effect of modifying the group sizes to be very imbalanced.
We investigate six different scenarios with varying group sizes but the same overall sample size of $n=250$ (see Table \ref{tab:ImbalancedMelon}). }
Figure \ref{fig:ImbalancedMelon} shows the mean misclassification rate over 50 repetitions. Scenario 1 and scenario 6, with the most extreme difference in the group sizes, result in the worst results for several methods. The LDA models are very stable but most of the time they are outperformed by LP. LP is only slightly affected by scenario 1 but otherwise it leads to similar results for the different settings outperforming all other methods.
Note that classification methods could be tuned in order to cope with imbalanced groups. \blue{For example, for random forests there are different strategies to adjust the group assignments if the group sizes are very different from each other \citep[e.g.][]{khoshgoftaar2007empirical, khalilia2011predicting}. However, according to the results shown in Figure  \ref{fig:ImbalancedMelon}, the performance of LP is very stable even in case of imbalanced groups.}

\iffigs
\begin{table}[!htb]
\centering
\caption{\blue{Group sizes for simulation scenarios for the Melon dataset. We vary the numbers of observations per group in order to simulate highly imbalanced group sizes.}}
 \label{tab:ImbalancedMelon}
 \begin{tabular}{lrrrrrr}
 \hline
   Scenario & 1 & 2 & 3 & 4 & 5 & 6\\
    \hline
  Class 1 & 25 & 50 & 75 & 100 & 125 & 150  \\
  Class 2 & 75 & 75 & 75 & 75 & 75 & 75 \\
  Class 3 & 150 & 125 & 100 & 75 & 50 & 25\\
   \hline
 \end{tabular}
 \end{table}
 \fi

\iffigs
\begin{figure}[!htb]
\centering
\includegraphics[width=\linewidth]{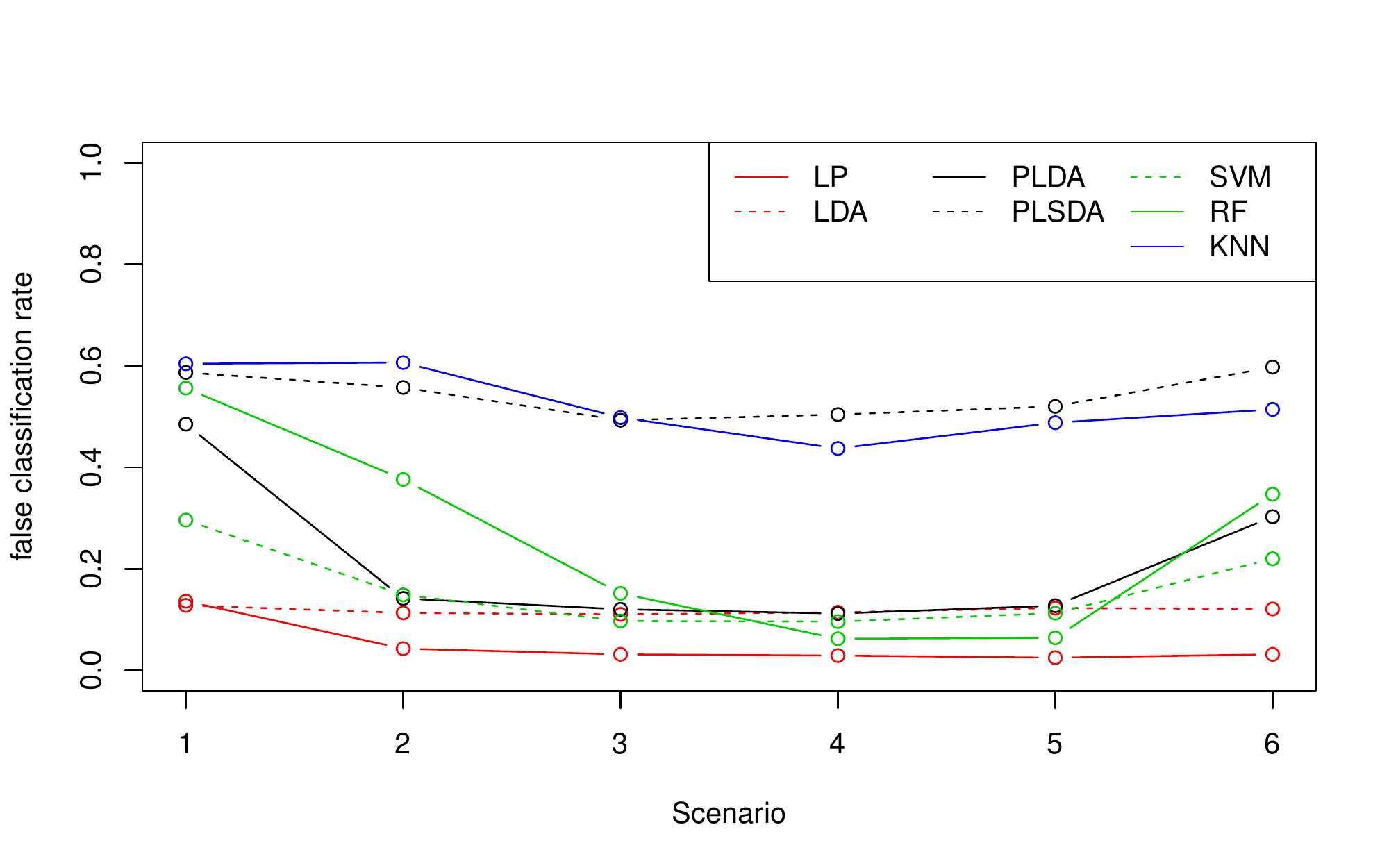} 
\caption{\blue{The performance of the evaluated classification methods for highly imbalanced datasets is evaluated. The six scenarios refer to the scenarios described in Table \ref{tab:ImbalancedMelon}. Especially setup 1 and 6 cause a problem for most approaches while LP presents itself as mostly robust towards imbalanced group sizes.}}
\label{fig:ImbalancedMelon}
\end{figure}
\fi

\section{Conclusions \& Outlook}\label{sec:conclusion}

\blue{We proposed a methodology for supervised classification combining aspects from the field of computer science and from the field of statistics. We use the concept of local projections to compute a set of linear discriminant models taking the information within each projection space and the distance to the projection spaces into account. The LDA models are then aggregated based on the projection-based degree of separation. As shown in \cite{ortner2017guided}, local projections can help to identify group structure in high-dimensional spaces. Therefore, this way of computing aggregated probabilities for class-membership allows the utilization of LDA for high-dimensional spaces while exploiting the advantages of identifying group structure by local projections.}

\blue{Additionally, a novel visualization based on ternary diagrams has been proposed which reveals links between the groups in high-dimensional space. The visualization makes use of the posterior probabilities computed from the local projections and therefore it allows to draw conclusions about the uncertainty of the class assignment supported by gray areas in the plot for uncertain assignment.}

\blue{The conducted evaluations on the performance of LP in comparison to related supervised classification methods (LDA, PLDA, PLSDA, SVM, RF and KNN) based on three different real-world datasets demonstrated the advantage of LP in}
\blue{various settings: two- and multi-group classification tasks, higher number of observations than variables and vice versa, inhomogeneous groups caused by outliers, and imbalanced group sizes. The only tuning parameter required for LP is the number $k$ of nearest neighbors, for which a lower and upper boundary has been proposed.}

\blue{While we utilize linear discriminant analysis performed on the projection space of each local projection, there is no reason to limit ourselves to LDA. Depending on the data setup, other methods can be preferred over LDA and still benefit from the local projection based aggregation. A general combination of classification approaches with local projections is still to be evaluated in future work.}

\bibliographystyle{apalike}

\end{document}